\documentclass[prd,aps,showpacs,preprintnumbers,amssymb]{revtex4}
\usepackage{axodraw}
\usepackage{color}
\usepackage{epsf}

\def\e3p{$\eta \rightarrow 3 \pi$}

\begin{document}
\title{%
\hfill{\normalsize\vbox{%
\hbox{}
 }}\\
{A top condensate model with a Higgs doublet and a Higgs triplet}}

\author{Renata Jora
$^{\it \bf a}$~\footnote[2]{Email:
 rjora@theory.nipne.ro}}
\author{Salah Nasri$^{\it \bf b}$~\footnote[3]{Email:
 nasri.salah@gmail.com}}
\author{Joseph Schechter
 $^{\it \bf c}$~\footnote[4]{Email:
 schechte@phy.syr.edu}}

\affiliation{$^{\bf \it b}$ National Institute of Physics and Nuclear Engineering PO Box MG-6, Bucharest-Magurele, Romania}
\affiliation{$^{\bf \it c}$ Department of Physics, College of Science, United Arab Emirates University, Al-Ain, UAE}

\affiliation{$^ {\bf \it d}$ Department of Physics,
 Syracuse University, Syracuse, NY 13244-1130, USA}

\date{\today}

\begin{abstract}
We reconsider top condensate models from the perspective that not only two quark composite states can form but also four quark ones. We obtain a model which contains
a Higgs doublet and  a Higgs triplet, where one of the neutral components of the Higgs triplet identifies with the Higgs boson found at the LHC. We discuss some of the phenomenological consequences.
\end{abstract}
\maketitle

\section{Introduction}

There are two mechanisms for spontaneous symmetry breaking of a gauge theory: the first one is through elementary scalars which develop vacuum expectation values; the second is through the formation of fermion condensates with the right quantum numbers, the dynamical symmetry breaking. The simplest example of dynamical symmetry breaking is furnished by the standard model itself.
We know that in the absence of quark masses and in the QCD presence the u, d, s (or only u, and d) are endowed with the chiral symmetry $SU(3)_L\times SU(3)_R$. The formation of vacuum condensates due to the strong group leads to the breaking of this group down to its vector part $SU(3)_V$. The corresponding Goldstone bosons, the pions couple with the correct quantum number to the gauge boson $W^{\pm}$ and Z thus providing them with the longitudinal degrees of freedom corresponding to massive particles. We know however that the scale of the condensates is 3000 times smaller that the electroweak scale and that pions although light are massive and present in experiments so they cannot play an actual role.
The simplest extension of this mechanism which might work for electroweak symmetry breaking  would be to consider  two quark composite states in the top, bottom sector.
The top quark condensate models have been discussed extensively in the literature \cite{Lindner}-\cite{Hill3} with or without the addition of new fermions or gauge symmetries.  The minimal top condensate theory where a regular Higgs is identified with top quark composite state leads to a too larger dynamical mass for the top quark without fine-tuning.  This value is determined from  the masses of the W and Z bosons in leading order  $\frac{1}{N_C}$.  The dynamical theory might get improved through introduction of an additional technicolor sector \cite{Hill}.
In general there are various constraints to the particle spectrum and interaction coming from the electroweak precision data and also from the latest results from the LHC.

In this work we want to reconsider the basic ideas of the top condensate models from the point of view largely employed in the light quark sector, that not only two quark but also four quark composite states can form. This would lead to additional Higgs multiplets which might cure the problems existent in the early versions of these models. We will not attempt to present a complete model together with it phenomenological analysis but rather to sketch a new theory and study some of its possible consequences.

\section{Short review of the QCD analogue}
The  spectroscopy of low lying pseudoscalar and scalar mesons suggest that these states may be an admixture of two quark and four quark structures \cite{Jora}.  The masses and mixings of these fields can be
described by a generalized linear sigma model with two chiral nonets, one with a two quark composition, the other one with a four quark one.

The schematic realization of these states must display the chiral $SU(3)_L\times SU(3)_R$ symmetry. The two quark bound state is written as:
\begin{eqnarray}
M^b_a=(q_{bA})^{\dagger}\gamma_4\frac{1+\gamma_5}{2}q_{aA}.
\label{two776}
\end{eqnarray}

We use a representation with the following form for the  $\gamma$ matrices and the charge conjugation matrix:
\begin{eqnarray}
\gamma_i=
\left[
\begin{array}{cc}
0&-i\sigma_i\\
i\sigma_i&0
\end{array}
\right],
\hspace{1cm}
\gamma_4=
\left[
\begin{array}{cc}
0&1\\
1&0
\end{array}
\right],
\hspace{1cm}
\gamma_5=
\left[
\begin{array}{cc}
1&0\\
0&-1
\end{array}
\right],
\hspace{1cm}
C=
\left[
\begin{array}{cc}
-\sigma_2&0\\
0&\sigma_2
\end{array}
\right].
\label{res44343}
\end{eqnarray}

With respect to the group $SU(3)_L\times SU(3)_R\times SU(3)_C$  the matrix M is in the $(3,{\bar 3},1)$ representation.

There are three possibilities for the four quark structures. The first one is that the four quark states are molecules made out of two quark-antiquark fields:
\begin{eqnarray}
M^{(2)b}_a=\epsilon_{acd}\epsilon^{bef}(M^{\dagger})_e^c(M^{\dagger})_f^d
\label{four6657}
\end{eqnarray}

Another possibility is that the four quark structures may be bound states of a diquark and anti-diquark. Here there are two choices.
In the first case the diquark is in ${\bar 3}$ of flavor, ${\bar 3}$ of color and has spin zero:
\begin{eqnarray}
&&L^{gE}=\epsilon^{gab}\epsilon^{EAB}q^T_{aA}C^{-1}\frac{1+\gamma_5}{2}q_{bB}
\nonumber\\
&&R^{gE}=\epsilon^{gab}\epsilon^{EAB}q^T_{aA}C^{-1}\frac{1-\gamma_5}{2}q_{bB}
\label{repr45554}
\end{eqnarray}

The matrix M has the structure:

\begin{eqnarray}
M_g^{(3)f}=(L^{gA})^{\dagger}R^{fA}
\label{res4443}
\end{eqnarray}

In the second case the diquark is in ${\bar 3}$ of flavor, 6 of color and has spin 1:
\begin{eqnarray}
&&L^g_{\mu\nu,AB}=L^g_{\mu\nu,BA}=\epsilon^{gab}Q^T_{aA}C^{-1}\sigma_{\mu\nu}\frac{1+\gamma_5}{2}Q_{bB}
\nonumber\\
&&R^g_{\mu\nu, AB}=R^g_{\mu\nu,BA}=\epsilon^{gab}Q^T_{aA}C^{-1}\sigma_{\mu\nu}\frac{1-\gamma_5}{2}Q_{bB}.
\label{res5554546}
\end{eqnarray}

Here $\sigma_{\mu\nu}=\frac{1}{2i}[\gamma_{\mu},\gamma_{\nu}]$. The matrix M has the form:
\begin{eqnarray}
M_g^{(4)f}=(L^g_{\mu\nu,AB})^{\dagger}R^f_{\mu\nu,AB}
\label{res5553443}
\end{eqnarray}

It can be shown using Fierz transformations that the three four quark structures are actually linearly dependent.

\section{Alternative scenarios}

Topcolor theories are based on extra strong gauge groups which act differently on the third generation. We will consider unitary groups  U(N) or SU(N) with $N\leq3$. These groups can be right handed, left handed or even vector groups.
Our purpose here  is to realize dynamical symmetry breaking with the minimal content of fermions but with additional gauge groups. For that, inspired by the low energy QCD
we will consider here the possibility that not only two quark but also four quark composite particles or condensates can form. We will start with the standard topcolor set-up which has
an extended $SU(3)_1\times SU(3)_2$ group acting on the top, bottom pair. Eventually this group should be broken down to $SU(3)_c$ to agree with the standard model.
This process limits the possible representation of the left handed and righthanded quarks.

We will consider here four scenarios which can lead to the correct color structure:

Scenario I:
\begin{eqnarray}
&&(t_L,b_L)\,\, {\rm are \,in} \,\,(3,1) \,\,{\rm of}\,\, SU(3)_1 \times SU(3)_2
\nonumber\\
&&(t_R,b_R)\,\, {\rm are\, in} \,\, (1,3) \,\,{\rm of}\,\, SU(3)_1 \times SU(3)_2.
\label{sc1}
\end{eqnarray}

 Scenario II:
\begin{eqnarray}
&&(t_L,b_R)\,\, {\rm  are\, in} \,\,(3,1)\,\, {\rm of}\,\, SU(3)_1 \times SU(3)_2
\nonumber\\
&&(t_R,b_L)\,\,{\rm are\,in}\,\, (1,3)\,\, {\rm of}\,\, SU(3)_1 \times SU(3)_2.
\label{sc2}
\end{eqnarray}

 Scenario III:
\begin{eqnarray}
&&(t_L, t_R, b_L)\,\, {\rm are\, in}\,\, (3,1) \,\,{\rm of}\,\, SU(3)_1 \times SU(3)_2
\nonumber\\
&&b_R\,\, {\rm is\, in} \,\,(1,3)\,\, {\rm of}\,\, SU(3)_1 \times SU(3)_2.
\label{sc3}
\end{eqnarray}

 Scenario IV:
\begin{eqnarray}
(t_L, t_R, b_L, b_R)\,\,{\rm are\, in}\,\, (3,1)\,\, {\rm of}\,\, SU(3)_1 \times SU(3)_2.
\label{sc4}
\end{eqnarray}

We will discuss all scenarios from the point of view of possible composite states and condensates.

 Scenario I

In this case one needs to form composite particles and condensates out of two left handed or righthanded fields. The two quark states should have a structure of the type:
$t_L^TC^{-1}t_L$ (or $t_L^TC^{-1}b_L$)which transform as $3^*$ of the $SU(3)_1$ group so they are not singlets.

For the four quark states we will consider only the left handed states as the same reasoning can be applied to the right handed ones. Then the only possibility is:
\begin{eqnarray}
&&(t_L^TC^{-1}t_L)^{\dagger}t_L^TC^{-1}t_L \,\,\,\, Y=0\,\,\,Q=0
\nonumber\\
&&(t_L^TC^{-1}t_L)^{\dagger}t_L^TC^{-1}b_L \,\,\, \,Y=0\,\,\,Q=-1
\label{doubl665}
\end{eqnarray}

Scenario II

One can form out two two quark composite states:
\begin{eqnarray}
&&b_R^{\dagger}t_L\,\,\,Y=1\,\,\,Q=1
\nonumber\\
&&t_R^{\dagger}b_L\,\,\,Y=-1\,\,\,Q=-1
\label{doubl6665}
\end{eqnarray}

These states cannot lead to two quark condensates and it is not possible to construct a two quark electrically neutral state.
The four quark states are:
\begin{eqnarray}
&&(b_R^TC^{-1}b_R)^{\dagger}t_L^TC^{-1}t_L\,\,Y=2\,\,Q=2
\nonumber\\
&&(t_R^TC^{-1}t_R)^{\dagger}t_L^TC^{-1}t_L\,\,Y=-2\,\,Q=0
\label{four554}
\end{eqnarray}

and again have the wrong quantum numbers to break the electroweak symmetry.

Scenario III

In this case the correct composite states can form:
\begin{eqnarray}
&&t^{\dagger}_R t_L\,\,Y=-1\,\,Q=0
\nonumber\\
&&t^{\dagger}_R b_L\,\,Y=-1\,\,Q=-1
\label{res4443}
\end{eqnarray}

These composite states form with the strength of the stronger gauge group $SU(3)_1$ (with gauge coupling $h_1$). This has also the advantage that the quark composite state
$b^{\dagger}_Rb_L$ can form only after the breaking of the topcolor group $SU(3)_1\times SU(3)_2$ down to $SU(3)_C$ with the strength of the color group. This can justify the
large difference between the mass of the top quark and that of the bottom.
There are various four quark states that might appear:
\begin{eqnarray}
(Q_R^TC^{-1}Q_R)^{\dagger}Q_L^TC^{-1}Q_L
\label{fourq4443}
\end{eqnarray}

where the right handed states are the top quark and left handed states are either the bottom or the top quark. Some of the previous states which connect only left handed or right handed particles are also possible.

 Scenario IV

In this case  $SU(3)_1$ is a vector group acting similarly to QCD. The fermions do not couple with $SU(3)_2$ which plays role of spectator. We need other particles to interact with $SU(3)_1\times SU(3)_2$ such that to break it down to $SU(3)_C$. The two quark and four quark composite states that can form are identical to those of regular QCD. A detailed description of these can be found in \cite{Jora}.

One can conclude that it is impossible to find four quark Higgs like structures for a particular scenario if two quark Higgs like composite states do not exist. Thus as in standard top color theories \cite{Hill} a doublet of the type described in Eq. (\ref{fourq4443}) or its four quark counterpart should spontaneously break the electroweak theory.

We will analyze in more detail Scenario III and IV since these are  the only ones that actually might work. In order to satisfy the gauge anomalies one might consider Scenario III in the context proposed in \cite{Hill}.  In order to prevent the formation of a large dynamical mass Scenario IV should be considered associated with an extra U(1) gauge symmetry as in \cite{Hill2}. Finally both scenarios should lead to the formation of the same composite states presented below.

The role of  the Higgs doublet is played by the composite state:
\begin{eqnarray}
\Phi_1=
\left[
\begin{array}{c}
t^{\dagger}_R b_L\\
t^{\dagger}_R t_L
\end{array}
\right].
\label{doubl6665}
\end{eqnarray}

There is triplet of four quark composite states with the hypercharge $Y=2$:
\begin{eqnarray}
&&\chi^{++}=n b_L^{\dagger}t_R b_L^{\dagger}t_R
\nonumber\\
&&\chi^+=n b_L^{\dagger}t_Rt_L^{\dagger}t_R
\nonumber\\
&&\chi^0=n t_l^{\dagger}t_Rt^{\dagger}t_R
\label{trpl88796}
\end{eqnarray}

Here n is a normalization factor which furnishes the correct vacuum structure.

There is also a  Higgs triplet with hypercharge $Y=0$.
\begin{eqnarray}
&&\xi^0=n (t_R^{\dagger}t_L)(t_R^{\dagger}t_L)^{\dagger}
\nonumber\\
&&\xi^+=n (t_R^{\dagger}t_L)(t_R^{\dagger}b_L)^{\dagger}
\nonumber\\
&&\xi^-=n (t_R^{\dagger}b_L)(t_R^{\dagger}t_L)^{\dagger}.
\label{tr77564}
\end{eqnarray}

The full Higgs triplet then takes the standard form \cite{Georgi}:
\begin{eqnarray}
\chi=
\left[
\begin{array}{ccc}
\chi^0&\xi^+&\chi^{++}\\
\chi^{-}&\xi^0&\chi^+\\
\chi^{--}&\xi^{-}&\chi^0
\end{array}
\right]
\label{full6575}
\end{eqnarray}

Thus the model contains a Higgs doublet and Higgs triplet.
The fields  in the model develop the vev's:
\begin{eqnarray}
&&\langle\Phi_0\rangle=\frac{a}{\sqrt{2}}
\nonumber\\
&&\langle\chi^0\rangle=b
\nonumber\\
&&\langle\xi^0\rangle=b
\label{veve443535}
\end{eqnarray}

Note that in this case the vacuum structure of the Higgs triplet is naturally,
\begin{eqnarray}
\langle\chi\rangle=
\left[
\begin{array}{ccc}
b&0&0\\
0&b&0\\
0&0&b
\end{array}
\right],
\label{veve4443}
\end{eqnarray}

as only one four quark vacuum condensate forms. Such a triplet can be used to generate non zero Majorana neutrino masses \cite{SV}, \cite{GR}.

The following relations hold:
\begin{eqnarray}
&&v^2=a^2+8b^2
\nonumber\\
&&c_H=\frac{a}{a^2+8b^2}
\nonumber\\
&&s_H=[\frac{8b^2}{a^2+8b^2}]^{1/2}.
\label{rel7756454}
\end{eqnarray}

After spontaneous symmetry breakdown the model contains the charged states (according to the classification under the SU(2) custodial symmetry \cite{Georgi}):
\begin{eqnarray}
&&H_5^{++}=\chi^{++}
\nonumber\\
&&H_5^+=\frac{1}{\sqrt{2}}(\chi^+-\xi^+)
\nonumber\\
&&H_3^+=\frac{a(\chi^{+}+\xi^+)-4b\Phi^+}{\sqrt{2}(a^2+8b^2)^{1/2}}
\label{st44335}
\end{eqnarray}

and the neutral states:
\begin{eqnarray}
&&H_5^0=\frac{1}{\sqrt{6}}(2\xi^0-\chi^0-\chi^{0*})
\nonumber\\
&&H_3^0=\frac{a(\chi^0-\chi^{)*})-2\sqrt{2}b(\Phi^0-\Phi^{0*})}{\sqrt{2}(a^2+8b^2)^{1/2}}
\nonumber\\
&&H_1=\frac{1}{\sqrt{2}}(\Phi_0+\Phi_0*)
\nonumber\\
&&H_1'=\frac{1}{\sqrt{3}}(\chi^0+\chi^{0*}+\xi^0).
\label{st4432}
\end{eqnarray}

\section{The dynamical sector}

Let us assume that below some scale one can integrate out massive gauge bosons, which might be the colorons or not to obtain four fermion interaction described by the term \cite{Lindner}:
\begin{eqnarray}
{\cal L }=G({\bar \Psi}_L^{iA}t_{RA})({\bar t}_R^{B}\Psi_{LiB})
\label{four6665}
\end{eqnarray}

Here i refers to $SU(2)_L$ indices whereas A, B are color indices.

If there is a strong sector which leads to the formation of top quark condensates one might consider the solution to the gap equation  for the top quark mass:
\begin{eqnarray}
&&m_t=-\frac{1}{2}G\langle{\bar t}t\rangle=
\nonumber\\
&&=2GN_c m_t\frac{i}{(2\pi)^4}\int d^4 p\frac{1}{p^2-m_t^2}.
\label{gappp776}
\end{eqnarray}

By analogy with QCD it is possible that also four quark composite states form. These can contribute to the dynamical mass of the top quark.
Then the solution to the gap equation becomes (see Fig. 0):
\begin{eqnarray}
&&m_t=\frac{G}{2}4N_c\frac{im_t}{(2\pi)4}\int d^4p \frac{1}{p^2-m_t^2}+
\nonumber\\
&&\frac{G^2}{4}\frac{i}{m_t}[(4N_c)(im_t)\int d^4p \frac{1}{p^2-m_t^2}]^2=
\nonumber\\
&=&m_t[2GN_c\frac{i}{(2\pi)^4}\int d^4p\frac{1}{p^2-m_t^2}+(2GN_c\frac{i}{(2\pi)^4}\int d^4p\frac{1}{p^2-m_t^2})^2]
\label{gap66565}
\end{eqnarray}

\begin{center}
\begin{picture}(300,100)(0,0)
\LongArrow(0,20)(30,20)
\LongArrow(60,20)(30,20)
\Text(80,20)[1]{=}
\Line(100,20)(160,20)
\Text(180,20)[1]{+}
\Line(200,20)(260,20)
\CCirc(130,30){10}{1}{White}
\CCirc(220,30){10}{1}{White}
\CCirc(240,10){10}{1}{White}
\Vertex(130,20){1}
\Vertex(220,20){1}
\Vertex(240,20){1}
\end{picture}\\
{\sl Fig.0. Contributions to the top mass}
\end{center}

We denote,
\begin{eqnarray}
2GN_c\frac{i}{(2\pi^4)}\int d^4p \frac{1}{p^2-m_t^2}=x
\label{den66578}
\end{eqnarray}

where x is a solution of the equation:
\begin{eqnarray}
1=x+x^2
\label{eq44232}
\end{eqnarray}

so it can have the values $\frac{-1-\sqrt{5}}{2}$ or $\frac{-1+\sqrt{5}}{2}$.

Solving Eq. (\ref{gap66565}) leads to :

\begin{eqnarray}
G=x[\frac{ N_c} {8\pi^2}[\Lambda^2-m_t^2\ln{\frac{\Lambda^2}{m_t^2}}]]^{-1}.
\label{res562438}
\end{eqnarray}

In the absence of the four quark condensate the sum of four fermions bubbles in the scalar channel leads to a scalar pole at $p^2=4m_t^2$ in accordance to the
standard results of the Nambu-Jona Lasinio model.
We will make a short analysis of the case where  two fermion or four fermions scalars might exist. We neglect the six fermions interaction since for scales higher than $m_t$ they
do not intervene anyway. Then the sum of the bubbles in the scalar channnel is given by \cite{Lindner}:

\begin{eqnarray}
&&\Gamma_s (p^2)=-\frac{G}{2}[1-2GN_c\frac{i}{(2\pi)^4}\int d^4 p\frac{1}{p^2-m_t^2}-
\nonumber\\
&&-G N_c(4m_t^2-p^2)\frac{i}{(2\pi)^4}\int d^4l\frac{1}{l^2-m_t^2}\frac{1}{(p+l)^2-m_t^2}]=
\nonumber\\
&&-\frac{G}{2}[x^2-G N_c(4m_t^2-p^2)\frac{i}{(2\pi)^4}\int d^4l\frac{1}{l^2-m_t^2}\frac{1}{(p+l)^2-m_t^2}]
\label{res4423131}
\end{eqnarray}

 To find the new pole one needs the solutions of the equation:
\begin{eqnarray}
N_c G(4m_t^2-p^2)\frac{1}{16\pi^2}\int_0^1 dz \ln[\frac{\Lambda^2-z(1-z)p^2}{m_t^2-z(1-z)p^2}]=x^2.
\label{sol88677}
\end{eqnarray}

It turns out that for $\Lambda\gg m_t$ the pole at $p^2=4m_t^2$ does not get removed as it is expected. This pole correspond to a two quark composite state. In order to see the four quark composite state one should consider a $\Lambda$  of order $m_t^2$ case in which  the new position of the pole is at $125-126$ GeV. Despite the fact that the actual masses will get modified at a more thorough analysis \cite{Lindner} the point of view we shall adopt in the rest of this work is that the two quark composite state corresponds to a heavier Higgs boson part of a Higgs doublet and the Higgs like particle discovered at the LHC corresponds to the four quark composite state which is part of a Higgs triplet.

\section{Relation between the dynamical sector and a Higgs model}

 The presence of the four fermion composite states with the quantum number of a Higgs triplet indicates that the dynamical model is well described by a standard model with a Higgs doublet and a Higgs triplet.
 Models which contain a Higgs doublet and a Higgs triplet have been discussed in the literature \cite{Georgi}-\cite{Moretti} both in connection to the LHC data and a possible Majorana mass term for the neutrinos. Most often the neutral component of the Higgs doublet or a combination of the neutral component of the Higgs doublet and the Higgs triplet have been identified with the Higgs boson found at the LHC. Our present work requires that a neutral component of the Higgs triplet identifies with the Higgs boson with a mass of $125-126$ GeV. We will discuss some of the implications of our scheme in connection with the vacuum  expectations values and of the top couplings. For simplicity we will assume that there is no mixing between the Higgs doublet and the Higgs triplet.

  As the Higgs triplet cannot couple at tree level with the quark fields it seems that we should not have significant contributions to the top quark mass (or the bottom one) from the four quark composite fields. However the dynamical mechanism suggests otherwise.

The gap equation suggest that the mass of the top quark is given by,
\begin{eqnarray}
m_t=m_tx+m_tx^2
\label{res55343}
\end{eqnarray}

where the first term corresponds to the two quark condensate contribution and the second to the four quark one.
We know that the dynamical model states that
\begin{eqnarray}
m_{t1}=xm_t=g_{t1}\langle H_1\rangle
\label{top887}
\end{eqnarray}

which suggests that we could also describe the second term by,
\begin{eqnarray}
m_{t2}=x^2m_t=g_{t2}\langle H_1'\rangle,
\label{four5546}
\end{eqnarray}

where $g_{t2}$ is an effective coupling for the Higgs singlet $H_1'$.
It is also known \cite{Lindner} that the four fermion interaction can be obtained by integrating out the two quark composite field through the interaction,

\begin{eqnarray}
L=L_{kin}+g_{t1}{\bar \Psi}_Lt_RH_1+h.c.
\label{int55454}
\end{eqnarray}

such that the relation,
\begin{eqnarray}
\frac{g_{t1}^2}{m_1^2}={\rm contribution\,to\,the\,four\,fermion\,interaction}
\label{rel8897}
\end{eqnarray}

holds.

However the Higgs triplet does not couple at tree level with the quarks but may couple through the interaction:
\begin{eqnarray}
L=L_{kin}+g_{t1}{\bar \Psi}_Lt_RH+\frac{g_{t}'}{M^3}{\bar \Psi}_Lt_R{\bar t}_R\Psi_L H'.
\label{copl887}
\end{eqnarray}

Here the third term contains generic contributions from the Higgs triplet corresponding to various hypercharges.
The term of interest for us  is that component of the Higgs triplet which upon spontaneous symmetry breaking develops a vev, that is $H_1'$.
Assuming that the third term in Eq. (\ref{copl887}) correspond to $H_1'$ we can deduce the effective coupling with the quark pair as being,
\begin{eqnarray}
\frac{g_t'}{M^3}\frac{B_1}{2}=g_{t2}.
\label{treelev554}
\end{eqnarray}

Here $B_1$ is the bubble corresponding to the two quark vacuum condensate:
\begin{eqnarray}
B_1=-im_t4N_c\int d^4p\frac{1}{p^2-m_t^2}.
\label{vac443}
\end{eqnarray}

Then the effective Higgs coupling for the triplet has exactly the same form as for the doublet and by integrating our $H_1'$ one obtains,
\begin{eqnarray}
\frac{g_{t2}^2}{m_2^2}+\frac{g_{t1}^2}{m_1^2}=G.
\label{rel99787655}
\end{eqnarray}

The following relations hold:
\begin{eqnarray}
&&\langle H_1\rangle=\frac{a}{\sqrt{2}}
\nonumber\\
&&\langle H_1'\rangle=\frac{3b}{\sqrt{3}}
\nonumber\\
&&a^2+8b^2=v^2
\label{res5554}
\end{eqnarray}

\section{Numerical estimates}

From Eqs. (\ref{top887}), (\ref{four5546}), (\ref{rel99787655}), (\ref{res5554}) and by taking as inputs the top quark mass and the $H_1'$ as the LHC boson with a mass $m_2=125.9$ GeV
one can obtain useful information about the vacuum structure, the Yukawa couplings, the mass of the neutral Higgs $H_1$ and the diphoton decay rate of $H_1'$.

Before proceeding we should mention that one of the strongest arguments against two  quark top condensate theories is that for low $\Lambda$ predict a too high mass for the top quark.
More exactly this constraint comes from a Pagels-Stokar formula which connects the composite Nambu-Goldstone bosons  to the top dynamical mass \cite{Lindner}
\begin{eqnarray}
\frac{v^2}{2}\approx\frac{N_c}{16\pi^2}m_t^2\int_0^1(1-x)\ln[\Lambda^2/[(1-x)m_t^2]],
\label{et554}
\end{eqnarray}
where  $v=246$ GeV for two quark composite models.

\begin{figure}
\begin{center}
\epsfxsize = 8cm
 \epsfbox{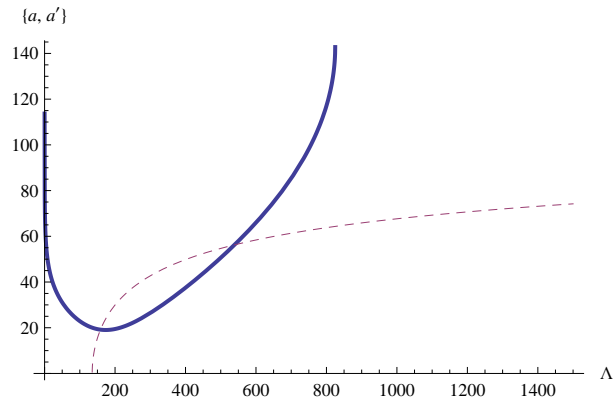}
\end{center}
\caption[]{%
Plot of the vacuum expectation values a (thick line) obtained for $m_1=347$ GeV  and a' (dashed line) as a function of the cut-off scale $\Lambda$.
}
\label{comp65}
\end{figure}

However in the case where also four quark composite states exist this relation is no longer true. Moreover assume we neglect in the r.h.s of Eq. (\ref{et554}) the four quark composite states then a similar relation should hold but with the l.h.s replaced by the vacuum expectation value of the Higgs doublet $a$.  One can compute this approximate value,
\begin{eqnarray}
a^{'2}=\frac{3}{16\pi^2}m_t^2[1+2\ln[\Lambda^2/m_t^2]]
\label{val887}
\end{eqnarray}

and  compare it to that obtained from Eqs. (\ref{top887}), (\ref{four5546}), (\ref{rel99787655}), (\ref{res5554}).  This is illustrated in Fig. \ref{comp65}. We will not consider this as an accurate estimate for a but rather as an argument that for low enough $a$ , $a\leq 80$ GeV one can obtain the correct mass for the top quark for reasonable values of $\Lambda$.

\begin{figure}
\begin{center}
\epsfxsize = 8cm
 \epsfbox{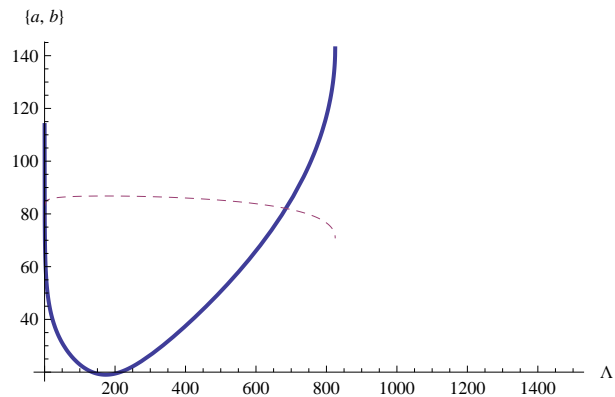}
\end{center}
\caption[]{%
Plot of the parameters a (thick line) and b (dashed line) as a function of $\Lambda$ for $m_1=347$ GeV.
}
\label{ft5545}
\end{figure}

\begin{figure}
\begin{center}
\epsfxsize = 8cm
 \epsfbox{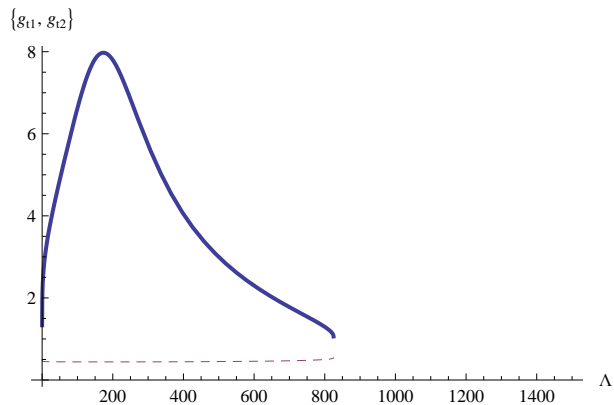}
\end{center}
\caption[]{%
Plot of the  parameters $g_{t1}$ (thick line) and $g_{t2}$ (dashed line) as a function of $\Lambda$ for $m_1=347$ GeV.
}
\label{t665}
\end{figure}

In Fig. \ref{ft5545} we plot the vacuum expectation values of the Higgs doublet, and of the Higgs triplet as computed from Eqs. (\ref{top887}), (\ref{four5546}), (\ref{rel99787655}), (\ref{res5554}) and for a mass of the neutral scalar $H_1$, $m_1=347$ GeV. Fig. \ref{t665} contains the effective couplings with the top quark pair of the Higgs bosons $H_1$ ($g_{t1}$)and $H_1'$ ($g_{t2}$).

\begin{figure}
\begin{center}
\epsfxsize = 8cm
 \epsfbox{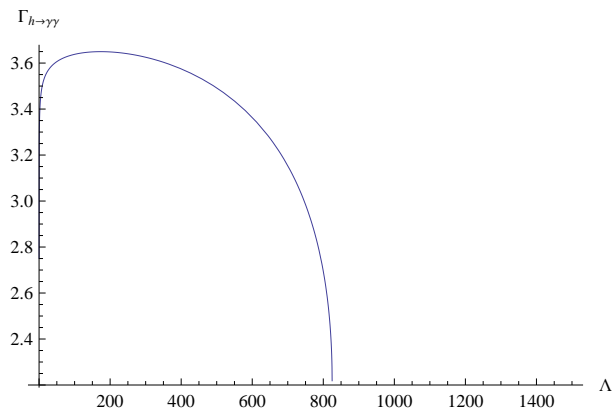}
\end{center}
\caption[]{%
Plot of the diphoton decay rate of the Higgs boson $H_1'$ as a function of the cut-off scale $\Lambda$.
}
\label{t543}
\end{figure}

We plot in Fig. \ref{t543} a gross estimate of the diphoton decay rate of the $126$ GeV Higgs boson, $H_1'$. Here we took into account only the charged W and top loops and neglected
completely the contribution of the other charged  Higgs bosons,  although the latter could be significant.  A calculation of the other Higgs bosons contribution would require an estimate of the masses of the these Higgs bosons in the actual model which is beyond the scope of the present work.

\begin{figure}
\begin{center}
\epsfxsize = 8cm
 \epsfbox{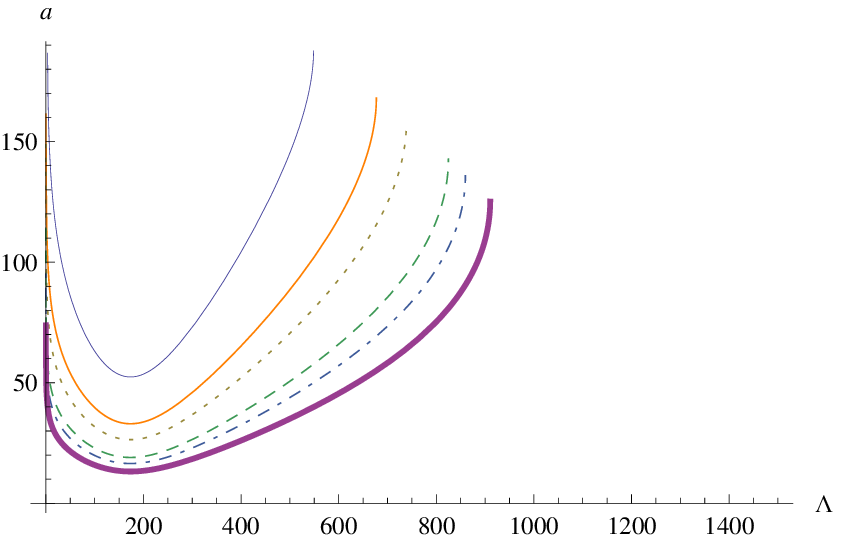}
\end{center}
\caption[]{%
Plot of the parameter a as a function of the cut-off scale $\Lambda$ for six values of the mass $m_1$ (GeV): $m_1=126$ (thin line), $m_1=200$ (orange line),
$m_1=250$ (dotted line), $m_1=347$ (dashed line), $m_1=400$ (dotdashed line), $m_1=500$ (thick line).
}
\label{t5438}
\end{figure}

\begin{figure}
\begin{center}
\epsfxsize = 8cm
 \epsfbox{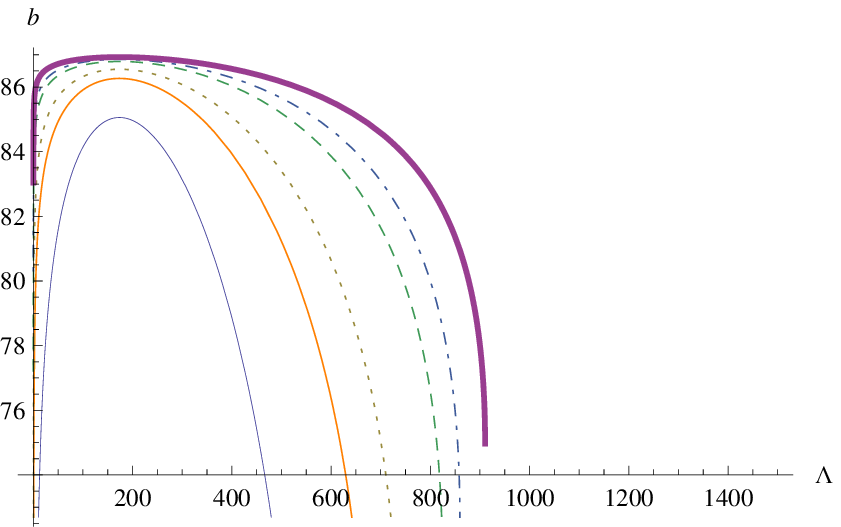}
\end{center}
\caption[]{%
Plot of the parameter b as a function of the cut-off scale $\Lambda$ for six values of the mass $m_1$ (GeV): $m_1=126$ (thin line), $m_1=200$ (orange line),
$m_1=250$ (dotted line), $m_1=347$ (dashed line), $m_1=400$ (dotdashed line), $m_1=500$ (thick line).
}
\label{t5433}
\end{figure}

We expect that a more thorough and more detailed calculation would modify the mass of the Higgs boson $H_1$ such that we extend our analysis to six possible masses of this particle
situated in the range $126-500$ GeV. We plot in Figs. \ref{t5438} and \ref{t5433} the vacuum expectation values of the Higgs doublet and triplet respectively for various masses of the Higgs boson $H_1$.

\begin{figure}
\begin{center}
\epsfxsize = 8cm
 \epsfbox{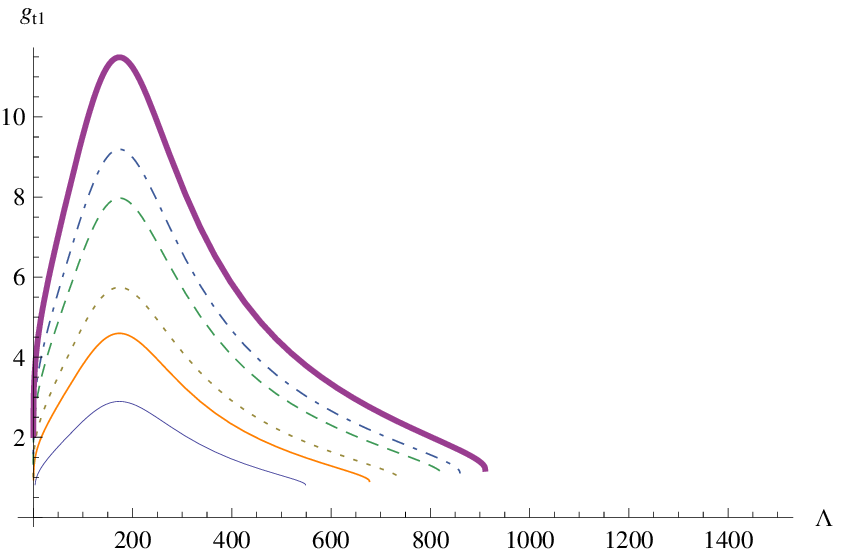}
\end{center}
\caption[]{%
Plot of the parameter $g_{t1}$ as a function of the cut-off scale $\Lambda$ for six values of the mass $m_1$ (GeV): $m_1=126$ (thin line), $m_1=200$ (orange line),
$m_1=250$ (dotted line), $m_1=347$ (dashed line), $m_1=400$ (dotdashed line), $m_1=500$ (thick line).
}
\label{t5434}
\end{figure}

\begin{figure}
\begin{center}
\epsfxsize = 8cm
 \epsfbox{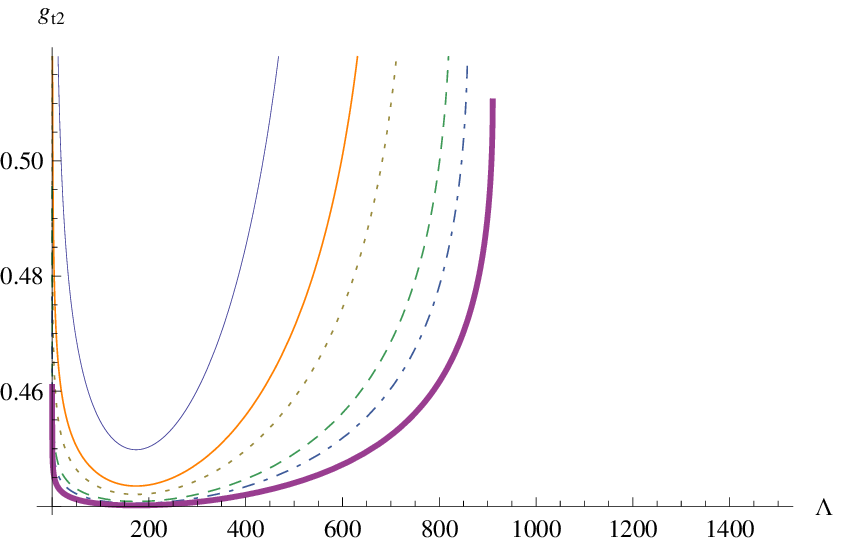}
\end{center}
\caption[]{%
Plot of the parameter $g_{t2}$ as a function of the cut-off scale $\Lambda$ for six values of the mass $m_1$ (GeV): $m_1=126$ (thin line), $m_1=200$ (orange line),
$m_1=250$ (dotted line), $m_1=347$ (dashed line), $m_1=400$ (dotdashed line), $m_1=500$ (thick line).
}
\label{t5435}
\end{figure}

\begin{figure}
\begin{center}
\epsfxsize = 8cm
 \epsfbox{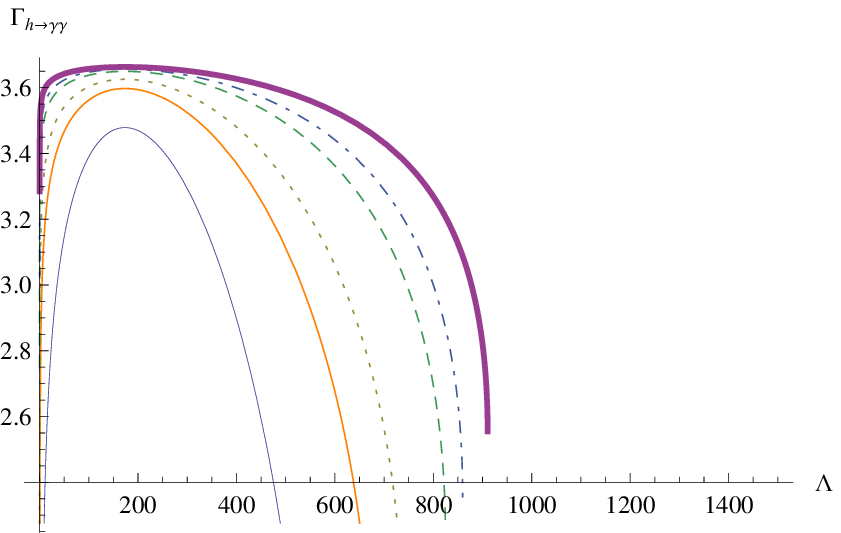}
\end{center}
\caption[]{%
Plot of the diphoton decay rate of the Higgs boson $H_1'$  as a function of the cut-off scale $\Lambda$ for six values of the mass $m_1$ (GeV): $m_1=126$ (thin line), $m_1=200$ (orange line),
$m_1=250$ (dotted line), $m_1=347$ (dashed line), $m_1=400$ (dotdashed line), $m_1=500$ (thick line).
}
\label{t5437}
\end{figure}

 Figs. \ref{t5434} and \ref{t5435} contain the effective top couplings for $H_1$ and $H_1'$ respectively as function of the cut-off scale $\Lambda$. For different masses of the Higgs boson $H_1$ we also plot the diphoton decay rate of the Higgs triplet component $H_1'$in Fig. \ref{t5437}.

One can conclude that large masses of the Higgs boson $H_1$ lead to a larger range for the allowed values of $\Lambda$.  For all masses considered the highest values of $\Lambda$ are
 in better agreement with the experimental value for the diphoton decay rate of the Higgs boson and also from the point of view of the preferred vacuum structure. In general the diphoton decay rate decreases with $m_1$ such that the lowest value is favored from this point of view.

\section{Conclusions}
 The most straightforward mechanism  of dynamical electroweak symmetry breaking is through the formation of top condensates.  We reevaluate the basic dynamical mechanism from the
 perspective that, as in the case of low-lying scalar mesons, not only two quark but also four quark composite states might exist.  Thus we arrive at an effective composite Higgs model
 which contains a Higgs doublet (two quark states)and a Higgs triplet (four quark states).  The underlying dynamics of the composite states suggests  that the Higgs boson found at the LHC should be identified to $H_1'$, the singlet component of the Higgs triplet. We showed that this theory can lead to a correct mass of the top quark for a reasonable value of the cut-off $\Lambda$.

We study some of the phenomenological consequences to find that the neutral component of the Higgs doublet $H_1$ can have a large range of masses and that the diphoton decay rate of the
Higgs boson $H_1'$  is naturally higher than the standard model predictions. The results could be improved in a full phenomenological analysis.

The purpose of this work was to sketch a new theory which might get improved by adding an extra technicolor sector etc. We delegate the problem of the other quark masses and couplings and of the possibility of flavor changing neutral currents in this context to further work.

\section*{Acknowledgments} \vskip -.5cm

The work of R. J. was supported by PN 09370102/2009. The work of J. S. was supported in part by the US DOE under Contract No. DE-FG-02-85ER 40231.

\end{document}